\documentclass[12pt]{article}
\usepackage{graphicx}        
\usepackage{latexsym}
\usepackage{amsmath}
\advance\textheight by \topskip
\begin{document}
\baselineskip 0.8cm
\thispagestyle{empty}
\begin{flushright}
KUCP0184\\
July 4, 2001\\ 
\end{flushright}
\vskip 2 cm
\begin{center}
{\LARGE \bf Thick Brane Worlds and Their Stability}
\vskip 1.7cm

{\bf Shinpei Kobayashi}
\footnote{E-mail: shinpei@phys.h.kyoto-u.ac.jp}
{\bf ,}
{\bf Kazuya Koyama}
\footnote{E-mail: kazuya@phys.h.kyoto-u.ac.jp}
{\bf and}
{\bf Jiro Soda}
\footnote{E-mail: jiro@phys.h.kyoto-u.ac.jp } \

\vskip 1.5mm

\vskip 2cm
 $^{1,2}$ Graduate School of Human and Environment Studies, Kyoto
University,
       Kyoto  606-8501, Japan \\
 $^3$ Department of Fundamental Sciences, FIHS, Kyoto University,
       Kyoto, 606-8501, Japan \\
\end{center}

\vskip 1cm
{\centerline{\large\bf Abstract}}
\begin{quotation}
\vskip -0.4cm
Three types of thick branes, i.e., Poincar\'{e}, de Sitter and  
Anti-de Sitter brane are considered. 
They are realized as the non-singular solutions of the Einstein 
equations with the non-trivial dilatons and the potentials. 
The scalar perturbations of these systems are also investigated.
We find that the effective potentials of the master equations 
of the scalar perturbations are positive definite and consequently 
these systems are stable under the small perturbations.

\end{quotation}

\newpage

\section{Introduction}

As the most promising candidate for the unified theory of everything, 
the superstring theory has been investigated for a long time. 
The recent discovery of D-branes stimulated a rather old idea, 
``brane world'' \cite{Rubakov1},\cite{BW2}.  
In particular, the Randall-Sundrum (RS) model has been advocated 
as a simple model \cite{RS1}, \cite{RS2}. 

RS considered the 3-brane (the four-dimensional Minkowski spacetime) 
embedded in the five-dimensional anti-de Sitter spacetime ($AdS_5$). 
And they found that there exist a massless graviton ($0$-mode) and 
massive gravitons (Kaluza-Klein modes). 
The massless graviton reproduces the Newtonian gravity on the 3-brane 
and Kaluza-Klein modes, which are the effect of the existence of 
the higher-dimension, give correction to the Newtonian gravity
\cite{GT},\cite{GKR}. 
RS showed that the Newtonian gravity can be reproduced in the
sufficiently low energy limit. 
And the idea that our universe is the 3-brane 
embedded in higher-dimensional 
spacetime came to be investigated eagerly. 
Furthermore, the cosmological consequences of the RS model have been 
investigated and there has been no contradiction with  
the observations until now \cite{S}-\cite{BWCP7}. 
The inflationary scenario also seems to be 
compatible with the brane models \cite{shinpei}-\cite{SHS}. 

Note that the RS model is motivated by the unified theory such as the 
superstring theory. Such theories are necessary   
to describe the high energy era of the history of our universe. 
And in the superstring theory, there seems to exist the minimum 
scale of the length, so we cannot consider an exact $0$-width brane. 
That is, we cannot neglect the thickness of the brane at the string 
scale.  

From these reason, ``thick brane world'' scenarios have been investigated 
\cite{DeWolfe}-\cite{Gremm2}.
In this paper, we especially pay attention to the construction 
of thick branes. We consider the three types of maximally symmetric 
branes, that is, Poincar\'{e}, 
de Sitter and anti-de Sitter brane. 
As these branes are highly-symmetric, they always exist as 
the solutions of the Einstein equations if we introduce 
non-trivial dilatons and suitable scalar potentials. 
And these solutions are non-singular in the whole spacetime. 
The thick de Sitter and the thick anti-de Sitter brane which are 
non-singular are found for the first time in this paper.

And we analyze the stability of these systems. 
We write down the master equations of the scalar 
perturbations and find the effective potential in the Poincar\'{e}, 
the de Sitter and the anti-de Sitter brane case, respectively. 
As a result, due to the positivity of the effective potentials, 
we find that all of systems are stable under the scalar perturbations. 

At last, we refer to the relation between the thickness and the 
non-commutativity. As mentioned above, the thickness is necessary  
due to the existence of the minimum length. And the non-commutativity 
also arises from the existence of the minimum length. So we can 
naively expect that the thickness has something to do with the 
non-commutativity.  

The organization of this paper is as follows: in Sec.2, 
we review the set-up 
of the thin brane models and construct three types of 
maximally symmetric branes. In order to see the effect of the 
thickness, we consider the behavior of gravitons in the thick brane 
backgrounds. In Sec.3 we analyze the scalar
perturbations of the thick brane systems. 
We derive the effective potentials of the scalar perturbations 
and examine the behavior of them. 
Sec.4 is devoted to the conclusions and the discussions. 
The relation between the thickness and the non-commutativity is also 
discussed there.

\section{Thick Brane Models}

\subsection{Thin Brane Models}

At first, we review the set-up of the thin brane models.
The simplest action of the thin brane model is 
\begin{equation}
S = \int d^5 x \sqrt{-g_5}\left(\frac{1}{2}R-\Lambda_5\right)
-\sigma \int d^4 x \sqrt{-g},
\end{equation}
where we used the unit $8\pi G_5 =1$ ($G_5$ is the five-dimensional 
gravitational constant), $g_5$ is the five-dimensional metric and 
$R$ denotes the five-dimensional Ricci scalar. If we consider the 
$AdS_5$, the five-dimensional cosmological constant $\Lambda_5$ is 
related to the AdS radius $l$ as
\begin{equation}
\Lambda_5 = -\frac{6}{l^2}.
\end{equation}
The second term of the action is the action of the brane and $\sigma$ 
denotes the tension of the brane.
Now we consider the following type of the metric,
\begin{equation}
ds^2 = dy^2 + e^{2\alpha(y)}\gamma_{\mu\nu}dx^{\mu}dx^{\nu},
\end{equation}
where $y$ denotes the coordinate of the direction of the bulk, 
$\gamma_{\mu\nu}$ is the 
metric on the brane, $\mu, \nu$ run the indices of the four-dimension 
of the brane. And  $e^{2\alpha(y)}$ is a so-called ``warp factor''.  
In this set-up, we can get three types of branes whose geometry 
are maximally symmetric, that is,  $M^4$, $dS_4$ or $AdS_4$. 
Solving the Einstein equations, we find  
\begin{eqnarray}
\alpha(y)= \begin{cases}
            y_0-|y|, &\mbox{Poincar\'{e} brane,} \\
            \log \left[ \sinh(y_0-|y|) \right], &\mbox{de Sitter brane,} \\
            \log \left[ \cosh(y_0-|y|) \right], &\mbox{Anti de Sitter brane},
                \end{cases}
\end{eqnarray}
where $y_0$ is a constant. 
In the case of the Poincar\'{e} brane, $y_0$ can be set to $0$ without the 
loss of the generality due to the Poincar\'{e} invariance. 
So we can say that $y_0$ does not have the physical meaning.  
In the case of the de Sitter brane, $y_0$ decides the range of the bulk. 
In fact, the warp factor becomes $0$ at $y=y_0$, so the range of $y$ 
is $-y_0 \leq y \leq y_0$ and $y=y_0$ is the horizon of the $AdS_5$.
On the contrary, in the case of the anti-de Sitter brane, 
$y_0$ denotes only the 
turning point of the warp factor because the metric does not 
become $0$ at $y=y_0$. So there is no horizon in $AdS_5$ with 
the anti-de Sitter slicing.
Now, we consider the extension of the thin brane systems to the
thick brane ones in the following subsections.

\subsection{Construction of the Thick Brane Models}

In order to realize thick brane models, we consider the following 
action,
\begin{equation}
S = \int d^5 x \sqrt{-g_5}\left[\frac{1}{2}R 
-\frac{1}{2}\left(\partial \varphi\right)^2 -V(\varphi)\right].
\end{equation}
Here $\varphi$ is the five-dimensional scalar field which depends only 
on the coordinate of the bulk and $V(\varphi)$ is its potential. 
We use the metric,
\begin{equation}
ds^2 = a^2(z)\left( dz^2 + \gamma_{\mu\nu}dx^{\mu}dx^{\nu}\right),
\end{equation}
where a conformal-like coordinate $z$ is defined through the following 
equations,
\begin{equation}
z \equiv \int \frac{dy}{a}, \hspace{1cm} a(z) = e^{\alpha(y(z))}.
\label{eq:y and z}
\end{equation}
Note that $\gamma_{\mu\nu}$ denotes the metric of 
the maximally symmetric four-dimensional spacetimes, so 
we can write 
the four-dimensional Ricci tensor and Ricci scalar as follows,
\begin{eqnarray}
 R^{(4)}_{\mu\nu} &=& 3K\gamma_{\mu\nu}, \\
 R^{(4)} &=& 12K,
\end{eqnarray}
where $K$ takes $0, 1$ or $-1$ and these values correspond 
Poincar\'{e}, de Sitter, Anti-de Sitter brane, 
respectively\footnote{
From now on, we also set AdS radius $l=1$ and we normalize the curvature
radius of the four-dimensional spacetime to the unity.}\ .  

Now we can write down the Einstein equations and the equation of motion 
of the scalar field (matter),
\begin{eqnarray}
\mbox{($z$,$z$)}&:& \hspace{0.5cm}
6\mathcal{H}^2 -6K = \frac{1}{2}\left(\varphi^{\prime}\right)^2 
                               -a^2 V(\varphi), 
\label{eq:zz-component of Einsten equation} \\
\mbox{($\mu$,$\nu$)}&:& \hspace{0.5cm}
3\mathcal{H}^{\prime}+3\mathcal{H}^2-3K 
= -\frac{1}{2}\left(\varphi^{\prime}\right)^2
                               -a^2 V(\varphi), 
\label{eq:munu-component of Einsten equation} \\
\mbox{matter}&:& \hspace{0.5cm}
\varphi^{\prime\prime}+3\mathcal{H}\varphi^{\prime} = 
                 a^2\frac{\partial V}{\partial \varphi}.
\label{eq:equation of matter}
\end{eqnarray}
Here a prime denotes the derivative with respect to $z$ and we defined  
$\mathcal{H}$ as follows,
\begin{equation}
\mathcal{H} \equiv \frac{a^{\prime}}{a}.
\end{equation}
From eqs.(\ref{eq:zz-component of Einsten equation}) and 
(\ref{eq:munu-component of Einsten equation}), we get the following 
equations,
\begin{eqnarray}
\left(\varphi^{\prime}\right)^2 &=& 
-3\mathcal{H}^{\prime}+3\mathcal{H}^2 -3K, 
\label{eq:equation of phi} \\
V(\varphi) &=& -\frac{1}{2a^2} 
\left(3\mathcal{H}^{\prime}+9\mathcal{H}^2 -9K \right).
\label{eq:equation of potential} 
\end{eqnarray}
From eqs.(\ref{eq:equation of phi}) and (\ref{eq:equation of potential}), 
we can see that one can construct a thick brane model
starting from a given warp factor $a(z)$, 
as long as $a(z)$ satisfies the condition,
\begin{equation}
(\varphi^{\prime})^2 \geq 0.
\label{eq:energy condition}
\end{equation}
Here the equation of motion of the scalar field 
(\ref{eq:equation of matter}) is automatically satisfied due to the 
Bianchi identity. Eq.(\ref{eq:equation of potential}) simply 
determines the functional form of the potential.

\subsection{Thick Poincar\'{e} Brane Model}

\begin{figure}[t]
\begin{center}
\includegraphics[width=8cm]{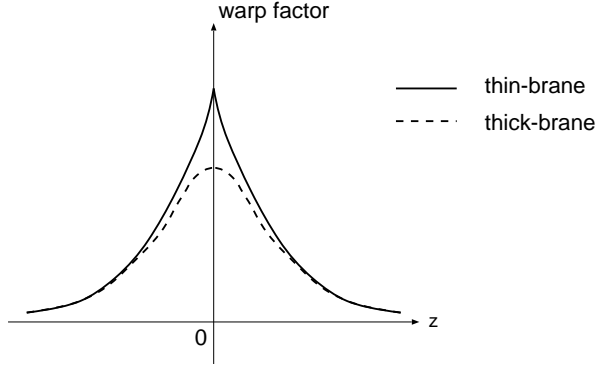}
\caption{The warp factors of the thin and the thick brane are shown. 
The solid line denotes the warp factor of the thin brane and 
the dashed line denotes that of the thick brane.}
\label{warp_factor of flat brane}
\end{center}
\end{figure}

In this paper, we use the following warp factor in order to make a thick  
Poincar\'{e} brane,\footnote{
There may be many ways to introduce the ``thickness''. Here we examine 
one of the possibilities. The parameter of the thickness which is 
related to the non-commutative geometry is discussed in Sec.4.}\ .
\begin{equation}
a^2(z) = e^{2\alpha(y(z))} = 
\left( \frac{1}{e^{-2n(y_0+y)}+e^{-2n(y_0-y)}} \right)^{1/n}.
\label{eq:warp factor for flat}
\end{equation}
As shown in Fig.\ref{warp_factor of flat brane}, there is no jump of 
the value of the extrinsic curvature at $y=y_0$. 
When we take the limit $n \to \infty$, this warp factor approaches 
to $e^{2(y_0-|y|)}$, which is the warp factor in the RS model. 
So the parameter $n$ controls the ``thickness'' of the brane.

In this case, we can write down $\varphi(y)$ and $V(\varphi)$ 
explicitly.
In fact, substituting the warp factor (\ref{eq:warp factor for flat}) 
into eqs. 
(\ref{eq:equation of phi}) and (\ref{eq:equation of potential}), we get 
\begin{equation}
\varphi(y) = \pm \sqrt{\frac{6}{n}}\mbox{Arctan} \left(e^{2ny}\right),
\end{equation}
\begin{equation}
V(\varphi) = -6 +3(n+2)\sin^2 \left(\frac{\sqrt{6n}}{3}\varphi \right).
\end{equation}

Now, in order to see the  effect of the ``thickness'', let us consider 
the gravitational perturbation in this system. 
Here the graviton $h_{\mu\nu}$ is the tensor perturbation of the 
metric, so it can be written as 
\begin{equation}
ds^2 = a^2(z)\left[ dz^2 + 
 (\gamma_{\mu\nu}+h_{\mu\nu})dx^{\mu}dx^{\nu}\right].
\end{equation}
Here $h_{\mu\nu}$ satisfies the transverse-traceless condition, 
\begin{equation}
h_{\mu}^{\mbox{ }\mu} = h_{\mu\nu}^{\mbox{ }\mbox{ }|\nu} = 0, 
\end{equation}
where the vertical bar denotes the covariant derivative with respect to 
$\gamma_{\mu\nu}$.
In this system, the equation for $h_{\mu\nu}$ becomes
\begin{equation}
h_{\mu\nu}^{\prime\prime}+3\mathcal{H}h_{\mu\nu}^{\prime}
+h_{\mu\nu|\lambda}^{\mbox{ }\mbox{ }\mbox{ }\mbox{ }|\lambda}
-2Kh_{\mu\nu} = 0.
\label{eq:eom of h}
\end{equation}
Here we can define the four-dimensional momentum $p$ as follows, 
\begin{equation}
h_{\mu\nu|\lambda}^{\mbox{ }\mbox{ }\mbox{ }\mbox{ }|\lambda}
-2Kh_{\mu\nu} = p^2 h_{\mu\nu}. 
\end{equation}
Furthermore, we can decompose $h_{\mu\nu}$ using the 
polarization tensor $\varepsilon_{\mu\nu}$ which depends only on the 
four-dimensional coordinates $x^{\rho}$ as follows,
\begin{equation}
h_{\mu\nu} (z,x^{\rho}) = \varepsilon_{\mu\nu}(x^{\rho}) X(z),
\end{equation}
where $\varepsilon_{\mu\nu}$ satisfies the transverse-traceless condition,
\begin{equation}
\varepsilon_{\mu}^{\mbox{ }\mu} 
= \varepsilon_{\mu\nu}^{\mbox{ }\mbox{ }|\nu} = 0.
\end{equation}
Now, regardless of the value of $K$,  we can rewrite eq.(\ref{eq:eom of h}) 
as, 
\begin{equation}
X^{\prime\prime}+3\mathcal{H}X^{\prime} = -p^2 X.
\label{eq:eom of Chi}
\end{equation}
To see the behavior of $h_{\mu\nu}$, we transform this equation into the 
Schr\"{o}dinger-type equation and write down the effective potential. 

\begin{figure}[t]
\begin{center}
\includegraphics[width=8cm]{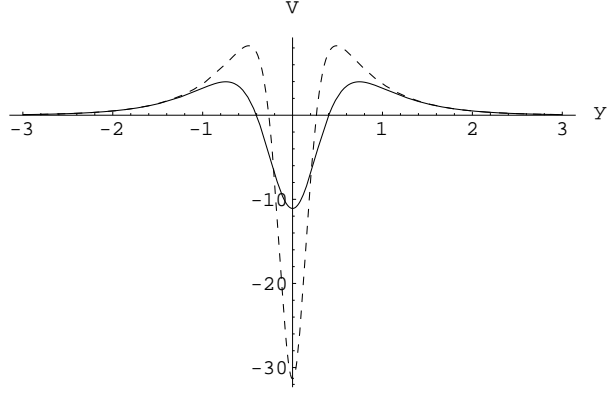}
\caption{The effective potential of the graviton in the 
thick Poincar\'{e} brane system. Here we set $y_0=1$. A solid line denotes
 $n=1$ case and a dashed line denotes $n=2$ case, respectively.}
\label{effe_pot_FreeScalar}
\end{center}
\end{figure}

Now we introduce a new function $\chi(z)$ which satisfies 
the following equation,
\begin{equation}
X(z) = a(z)^{-3/2} \chi(z).
\label{eq:chi}
\end{equation}
Substituting eq.(\ref{eq:chi}) into eq.(\ref{eq:eom of Chi}), 
we find that the equation for $\chi(z)$ becomes
\begin{equation}
-\chi^{\prime\prime}(z)+V(z)\cdot\chi(z) = -p^2 \chi(z),
\end{equation}
\begin{equation}
V(z) = \frac{3}{2}\mathcal{H}^{\prime}+\frac{9}{4}\mathcal{H}^2, 
\label{eq:effective potential of free scalar}
\end{equation}
where $V(z)$ is the effective potential of the graviton.
Furthermore, substituting the warp factor (\ref{eq:warp factor for flat}) 
into eq.(\ref{eq:effective potential of free scalar}), we get 
\begin{equation}
V(z) = \frac{3e^{2y_0}\left[5e^{4ny}-(16n+10)+5e^{-4ny}\right]}
         {4(e^{2ny}+e^{-2ny})^{2+\frac{1}{n}}}.
\label{eq:effective potential of free scalar2}
\end{equation}
We show the effective potentials in Fig.\ref{effe_pot_FreeScalar}. 
The effective potentials show that the gravitons can localize 
around the brane.
In fact, there is a $0$-mode solution,
\begin{equation}
\chi_0(z) \sim a^{3/2}(z). 
\label{eq:0-mode}
\end{equation}
Here we neglect the normalization constant.

The width of the brane becomes wider as $n$ approaches to $0$. 
On the other hand, when $n$ approaches to $\infty$, 
the effective potential approaches the 
``volcano'' potential which can be seen in the RS model. 
And the $0$-mode solution $\chi_0(z)$ coincides with the massless 
graviton in the RS model.  
So we can say that if we take a sufficiently large $n$ and the low
energy limit, the Newtonian gravity can be reproduced 
in this model\footnote{
In \cite{Csaki}, it is shown that the Newtonian gravity is reproduced. 
}\ .

\subsection{Thick De Sitter Brane Model}

\begin{figure}[t]
\begin{center}
\includegraphics[width=8cm]{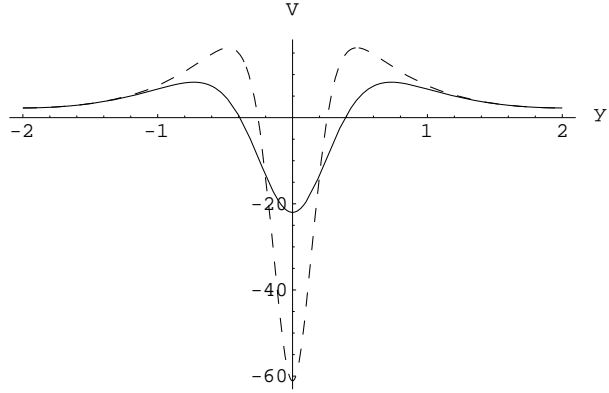}
\caption{The effective potential of the graviton in the thick de Sitter 
brane system. A solid line denotes $n=1$ case and a dashed line denotes 
$n=2$ case, respectively. Both approach to $\frac{9}{4}H^2$ as $y\to y_0$. 
Here we set $y_0=2$.}
\label{dSnograviton}
\end{center}
\end{figure}

Similarly, we set the warp factor of a thick de Sitter brane as follows,
\begin{equation}
a^2(z) = \left(\frac{1}{\sinh^{-2n}(y_0+y)
                     +\sinh^{-2n}(y_0-y)}\right)^{1/n}.
     \label{eq:thick_dS_warp_factor} 
\end{equation}
In this case, we cannot calculate $\varphi(y)$ and $V(\varphi)$ 
analytically. So we have to calculate them numerically.

Let us consider the behavior of the graviton in the thick de Sitter
brane system as well as in the previous subsection. Substituting 
the warp factor (\ref{eq:thick_dS_warp_factor}) into 
eq.(\ref{eq:effective potential of free scalar2}), we get the effective 
potential of the graviton propagating in the thick de Sitter brane 
system. The result is shown in Fig.\ref{dSnograviton}.

The shape of the effective potential in this case 
is very similar to that in the case of the Poincar\'{e} brane. 
There is a hole around the location of the brane
and we can find a bound state of the graviton which localize around the 
brane.

Here, we have to note that the asymptotic value 
of the effective potential in this case is different from that in the 
case of the Poincar\'{e} brane. 
The value of the effective potential of the graviton in the case of 
the thick de Sitter brane approaches to $\frac{9}{4}H^2$ as 
$y\to\pm\infty$ (where $H$ is the Hubble constant). 
This phenomenon can be seen in the analysis of the thin brane, too 
\cite{shinpei},\cite{GS}.  

\subsection{Thick Anti-De Sitter Brane Model}

\begin{figure}[t]
\begin{center}
\includegraphics[width=8cm]{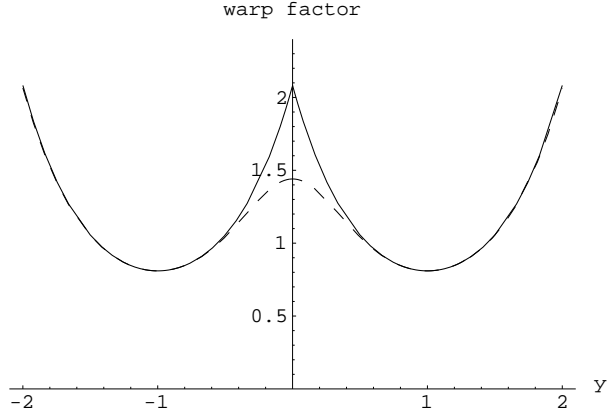}
\caption{The squares of a warp factors of a thin and thick AdS brane. 
A solid line denotes the warp factor of the thin AdS brane 
and a dashed line denotes that of the thick AdS brane. Here we set $y_0=1$.}
\label{AdSnowarp}
\end{center}
\end{figure}

\begin{figure}[t]
\begin{center}
\includegraphics[width=8cm]{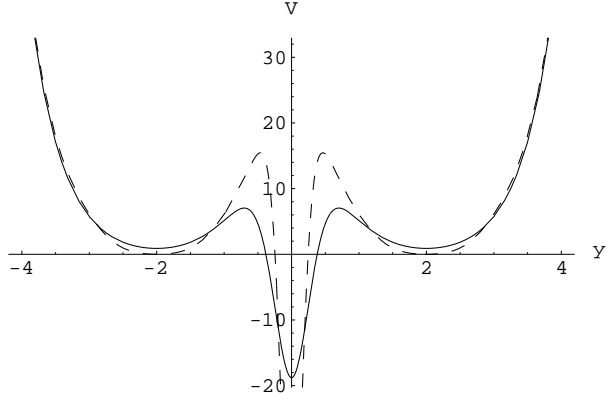}
\caption{The effective potential of the graviton in the thick AdS 
brane system. A solid line denotes $n=1$ case and a dashed line 
denotes $n=2$ case, respectively. Here we set $\epsilon = 0.4$ and $y_0=2$.}
\label{AdSnograviton}
\end{center}
\end{figure}

At last, we construct the warp factor of the thick anti-de Sitter brane 
as follows,
\begin{equation}
a^2(z) =\left[ \left( 1 +\frac{1}{\sinh^{-2n}(y_0+y)
+\sinh^{-2n}(y_0-y)}\right)^{1/2n} -\epsilon \right]^2,
\label{eq:thick_AdS_warp_factor}
\end{equation}
where $\epsilon$ is a constant which satisfies $0<\epsilon<1$. 
$\epsilon$ is necessary for the spacetime to become an exact $AdS_5$ 
only at infinity (i.e., $y=\pm \infty$). Without $\epsilon$, we find that 
the spacetime also becomes an exact $AdS_5$ at $y=\pm y_0$. 
This will cause a technical trouble in the analysis of the scalar perturbation 
discussed in Sec.3. And we have to calculate the fluctuations numerically 
in this case as well. We show the 
warp factor of the thick Anti-de Sitter brane in Fig.\ref{AdSnowarp} and 
the effective potential of the graviton propagating in this system in 
Fig.\ref{AdSnograviton}. Both of them diverge as $y \to \infty$, 
which is different from the previous two cases.

\section{Stability Analysis}

\subsection{Perturbation of Scalar-Gravity Coupled Systems}

In the previous section, we constructed three types of the thick brane 
systems. Next, we have to investigate the stability of them. To do so, 
let us examine the scalar perturbation of these systems. 

We use the following perturbed metric, 
\begin{eqnarray}
ds^2 &=& (g_{MN}+\delta g_{MN})dx^{M}dx^{N} \nonumber \\
&=& a^2(z)\big[(1+2\phi)dz^2-2B_{|\mu}dzdx^{\mu} 
+\left((1+2\psi)\gamma_{\mu\nu}-E_{|\mu\nu}\right)dx^{\mu}dx^{\nu}\big].
\end{eqnarray}
In this paper we use the longitudinal gauge ($B=E=0$) and we get 
the following metric,
\begin{equation}
ds^2 = a^2(z)\left[(1+2\phi)dz^2
+(1+2\psi)\gamma_{\mu\nu}dx^{\mu}dx^{\nu}\right].
\end{equation}
Now we get equations for the scalar perturbation, 
\begin{eqnarray}
\mbox{($z,z$)}&:& 
3\gamma^{\rho\lambda}\psi_{|\rho\lambda}+12\mathcal{H}\psi^{\prime}
-12\mathcal{H}^2\phi +12K\psi = \varphi_0^{\prime}\delta\varphi^{\prime} 
-\phi (\varphi_0^{\prime})^2 
-a^2 \frac{\partial V}{\partial \varphi_0}\delta\varphi,
\label{eq:perturbation-zz} \\
\mbox{($z,\mu$)}&:&
-3\psi_{|\mu}^{\prime}+3\mathcal{H}\phi_{|\mu} 
= \varphi_0^{\prime}\delta\varphi_{|\mu}, 
\label{eq:perturbation-zmu} \\
\mbox{($\mu,\nu$)}&:&
\Big(3\psi^{\prime\prime}-6\mathcal{H}^{\prime}\phi
-3\mathcal{H}\phi^{\prime}+9\mathcal{H}\psi^{\prime}
-6\mathcal{H}^2\phi 
+\gamma^{\rho\lambda}\phi_{|\rho\lambda}
+2\gamma^{\rho\lambda}\psi_{|\rho\lambda}+6K\psi\Big)
\delta_{\ \nu}^{\mu} \nonumber \\
& &\hspace*{2cm}-\gamma^{\mu\rho}\phi_{|\rho\nu}
-2\gamma^{\mu\rho}\psi_{|\rho\nu}
=\left(-\varphi_0^{\prime}\delta\varphi^{\prime}
+\phi(\varphi_0^{\prime})^2-a^2 
\frac{\partial V}{\partial \varphi_0}\delta\varphi \right)
\delta_{\ \nu}^{\mu},
\label{eq:perturbation-munu} \\
\mbox{matter}&:&
\delta\varphi^{\prime\prime}+3\mathcal{H}\delta\varphi{\prime}
+(4\psi^{\prime}-\phi^{\prime}-6\mathcal{H}\phi)\varphi_0^{\prime}
-2\phi\varphi^{\prime\prime}
+\gamma^{\rho\lambda}\delta\varphi_{|\rho\lambda}
=a^2\frac{\partial^2 V}{\partial \varphi_0^2}\delta\varphi,
\label{eq:perturbation-matter}
\end{eqnarray}
where the vertical bar denotes the covariant derivative with respect to 
$\gamma_{\mu\nu}$ just as in Sec.2. And we splitted $\varphi$ into 
its background $\varphi_0$ and its perturbations $\delta\varphi$.
Now let us derive the master equation of this system. 
At first, from eq.(\ref{eq:perturbation-zmu}), we get 
\begin{equation}
\delta\varphi = \frac{1}{\varphi_0^{\prime}}
\left(-3\psi^{\prime}+3\mathcal{H}\phi\right),
\label{eq:conclusion-perturbation-zmu}
\end{equation}
and from the off-diagonal part of eq.(\ref{eq:perturbation-munu}), we get 
\begin{equation}
\phi +2\psi = 0. 
\label{eq:perturbation-off-diagonal-munu}
\end{equation}
Substituting eqs.(\ref{eq:perturbation-matter}), 
(\ref{eq:conclusion-perturbation-zmu}) and  
(\ref{eq:perturbation-off-diagonal-munu}) into 
eq.(\ref{eq:perturbation-zz})$+$(\ref{eq:perturbation-munu}), we find 
the master equation of the system, 
\begin{equation}
\psi^{\prime\prime}+\gamma^{\rho\lambda}\psi_{|\rho\lambda}
+\left(3\mathcal{H}
-2\frac{\varphi_0^{\prime\prime}}{\varphi_0^{\prime}}\right)\psi^{\prime}
+\left(4\mathcal{H}^{\prime}-4\mathcal{H}
\frac{\varphi_0^{\prime\prime}}{\varphi_0^{\prime}}
+6K\right)
\psi=0.
\end{equation}

We transform this equation into the form of Schr\"{o}dinger equation 
in order to examine the stability of this system.
To do so, we define a new function $F(z,x^{\mu})$ as, 
\begin{equation}
\psi(z,x^{\mu})= \frac{\varphi_0^{\prime}(z)}{a(z)^{3/2}} F(z,x^{\mu}).
\end{equation}
We get the Schr\"{o}dinger-type equation for the scalar perturbation,
\begin{equation}
-F^{\prime\prime}(z, x^{\mu})+ V_{e}(z)\cdot F(z, x^{\mu}) =  
\gamma^{\rho\lambda}F(z, x^{\mu})_{|\rho\lambda},
\label{eq:Schrodinger equation}
\end{equation}
where $V_e$ is the effective potential for the perturbation and 
its concrete form becomes 
\begin{equation}
V_{e} = -\frac{5}{2}\mathcal{H}^{\prime}
+\frac{9}{4}\mathcal{H}^2 
+\mathcal{H}\frac{\varphi_0^{\prime\prime}}{\varphi_0^{\prime}}
-\frac{\varphi_0^{\prime\prime\prime}}{\varphi_0^{\prime}}
+2\left(\frac{\varphi_0^{\prime\prime}}{\varphi_0^{\prime}}\right)^2
-6K.
\label{eq:effective potential}
\end{equation}
We analyze the effective potentials of three types of branes, 
separately.

\subsection{Thick Poincar\'{e} Brane Case}

In the case of the thick Poincar\'{e} brane, we can expand $F(z,x^{\mu})$ 
as follows,
\begin{equation}
F(z,x^{\mu}) = \int \frac{d^4 p}{(\sqrt{2\pi})^4} 
f_p(z) e^{ip_{\mu}x^{\mu}},
\label{eq:expansion for flat}
\end{equation}
and we find that the equation for $f_p(z)$ becomes 
\begin{equation}
-f_p^{\prime\prime}(z)+ V_{e}(z)\cdot f_p(z)= m^2 f_p(z),
\end{equation}
where $m$ is the four-dimensional mass which satisfies $m^2=-p^2$. 
In the four-dimensional flat 
spacetime, we have the relation $p^2 = -\omega^2 + k^2$. Here, 
$\omega$ is 
the eigenvalue of the time-direction and $k$ is the norm of 
the three-dimensional momentum. If there is only time-dependence in 
this system (i.e., $k=0$), $\omega =m$. So if there is the solution 
which has an imaginary $m$, we can say that this system is unstable.  

Next we examine the effective potential $V_{e}$. In this system, $V_e$ 
becomes
\begin{equation}
V_{e} = \frac{e^{2y_0}\left[(16n^2+16n+3)e^{4ny}+32n^2+80n-6
                       +(16n^2+16n+3)e^{-4ny}\right]}
            {4(e^{2ny}+e^{-2ny})^{2+\frac{1}{n}}}.
\end{equation}
Clearly $V_{e}$ is positive definite and it approaches to $0$ as 
$y \to \pm \infty$.\footnote{Note that we can discuss using $y$ instead
of $z$ because $z$ is a monotonic function of $y$ from 
eq.(\ref{eq:y and z}).} \
The solution which has an imaginary $m$ cannot exist because it will 
necessarily diverge either at $y=\infty$ or at $y=-\infty$.  
So we can conclude that the thick Poincar\'{e} brane is stable under 
the scalar perturbation. 
It is interesting that there is no bound state on the brane, 
which is different from the 
tensor perturbation. 
The thickness parameter $n$ determines the height 
of the effective potential. The effective potential becomes higher as 
$n$ becomes larger. In the thin brane limit (i.e.,$n \to \infty $), 
the height of the effective potential get to infinity, and 
the thin brane becomes a singular object which has no width.

\begin{figure}[t]
\begin{center}
\includegraphics[width=8cm]{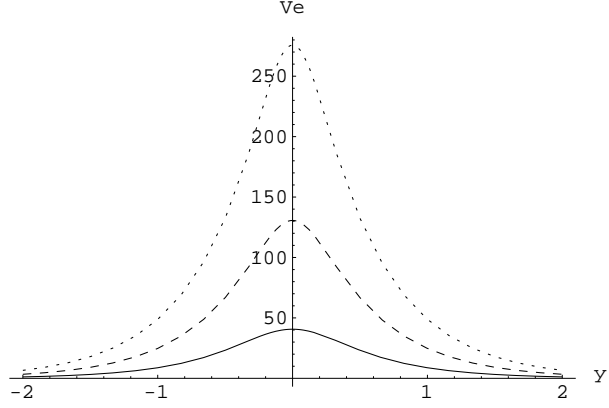}
\caption{The effective potential of the scalar perturbation in the case 
of the thick Poincar\'{e} brane. 
A solid line, a dashed line and a dotted line 
denotes $n=1$, $n=2$ and $n=3$ case, respectively.}
\label{effe_pot_of_flat_thick-brane}
\end{center}
\end{figure}

\subsection{Thick De Sitter Brane Case}

In the case of the thick de Sitter brane, we expand $F(z,x^{\mu})$ as 
follows, 
\begin{equation}
F(z,x^{\mu}) = \int d^3\mbox{\boldmath $k$}dm \mbox{ }
     f_m(z) g_{km}(t)e^{i\mbox{\boldmath $kx$}}.
\label{eq:expansion for de Sitter}
\end{equation}
Here we introduce the four-dimensional mass $m$ through 
the equation for $g_{km}(t)$ as follows, 
\begin{equation}
\ddot{g}_{km}(t)+3H\dot{g}_{km}(t)
+\left( \mbox{\boldmath $k$}^2 e^{-2t} +m^2 \right)g_{km}(t)=0,  
\label{eq:definition of mass}
\end{equation}
where a dot denotes the derivative with respect to $t$ and 
$H$ is the Hubble constant. 
Then solving eq.(\ref{eq:definition of mass}), we find that 
$g_{km}$ becomes
\begin{equation}
g_{km}(\eta) = \frac{\sqrt{\pi}}{2}H(-\eta)^{3/2}e^{-\pi\beta/2}
H_{i\beta}^{(1)}(-k\eta),
\end{equation}
where $H^{(1)}$ is the Hankel function of the first kind\footnote{
We set $H=1$ from now on.}\ . 
Here, $\eta$ is conformal time defined as 
\begin{equation}
\eta \equiv -e^{-t},
\label{eq:conformaltime}
\end{equation}
and $\beta$ is defined as follows,  
\begin{equation}
\beta \equiv \sqrt{m^2-\frac{9}{4}}.
\label{eq:beta}
\end{equation}
In order to investigate the time evolution of the scalar perturbation, 
we examine 
$g_{km}(\eta)$ for various $m$. At first, we expand $g_{km}(\eta)$ near 
$\eta \sim 0$ ($t \to \infty$) as 
\begin{eqnarray}
g_{km}(\eta) &=& \frac{\sqrt{\pi}}{2}e^{-\pi\beta/2}
\Big[ \frac{k^{i\beta}}{2^{i\beta}\Gamma(1+i\beta)}
\left(1+i\frac{\cos(i\beta\pi)}{\sin(i\beta\pi)}\right)
(-\eta)^{\frac{3}{2}+i\beta} \nonumber \\ 
& &\hspace*{4cm}-\frac{ik^{-i\beta}}{2^{-i\beta}\sin(i\beta\pi)
\Gamma(1-i\beta)}(-\eta)^{\frac{3}{2}-i\beta}\Big].
\label{eq:expansion of g_p}
\end{eqnarray}
For $m^2 \geq \frac{9}{4}$, $\beta$ becomes a real number, 
so $g_{km}$ oscillates near $\eta \sim 0$ and we can 
say that $g_{km}$ is stable. Next, for $0 < m^2 \leq \frac{9}{4}$, 
$\beta$ becomes an imaginary number. 
Now we define a new variable $\zeta$ as 
\begin{equation}
\beta = \sqrt{m^2-\frac{9}{4}} \equiv i \zeta.
\end{equation}
The first term in the bracket of eq.(\ref{eq:expansion of g_p}) 
behaves as $(-\eta)^{\frac{3}{2}-\zeta}$ and the second term  
behaves as $(-\eta)^{\frac{3}{2}+\zeta}$. So both terms converge to 
$0$ as $\eta$ approaches to $0$ because $\zeta$ is 
$0\leq \zeta < \frac{3}{2}$. 
From the above discussion, we can conclude that the solution whose 
$m^2$ satisfies  $0 < m^2 \leq \frac{9}{4}$ is also stable. 

At last, for $m^2 \leq 0$, $\zeta$ becomes 
larger than $\frac{3}{2}$, so the first term of 
eq.(\ref{eq:expansion of g_p}) does not converge to $0$ when $\eta$ gets 
to $0$. It means that if there is the solution with $m^2 \leq 0 $, 
this system must be unstable. 

\begin{figure}[t]
\begin{center}
\includegraphics[width=8cm]{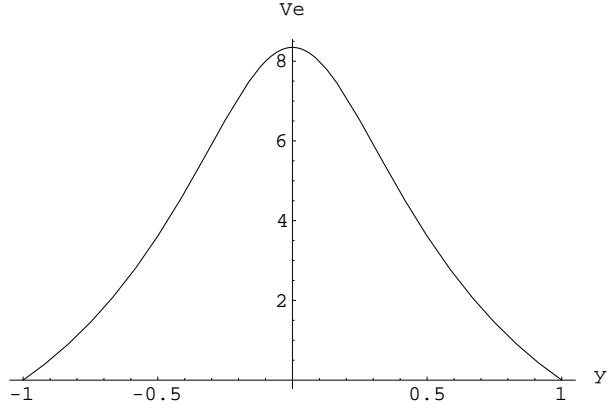}
\caption{The effective potential of the scalar perturbation in the 
case of the thick de Sitter brane. Here we set $n=1, y_0=1$.}
\label{effe_pot_of_dS_thick-brane}
\end{center}
\end{figure}

Then we examine the effective potential for the thick de Sitter brane. 
We calculate it numerically and the result is shown in 
Fig.\ref{effe_pot_of_dS_thick-brane}.
In this case, $y=\pm y_0$ is the horizon of $AdS_5$, so the spacetimes 
is defined in the region $-y_0 \leq y \leq y_0$. 
Fig.\ref{effe_pot_of_dS_thick-brane} shows that the effective 
potential of the scalar perturbation is positive 
for $-y_0 \leq y \leq y_0$ and we can conclude that there is no 
solution with $m^2 \leq 0$ because it cannot be normalizable. 
So unstable solutions do not exist, as a result, this system is 
stable under the scalar perturbation just as in the case of 
the thick Poincar\'{e} brane.

Notice that the shape of the 
effective potential is very different from that of the graviton or 
that of the free test scalar field. The effective potentials of the 
graviton or the free scalar field have holes in their potentials 
at the location of the branes. On the contrary, the effective potential 
of the scalar perturbation has no hole at the location of the brane. 
Furthermore, the mass gap has disappeared in the effective potential 
of the scalar perturbation. There exist the mass gaps in the effective 
potentials of the graviton or the free test scalar fields.
That is, the effective potentials approach to $\frac{9}{4}H^2$ 
($H$ is the Hubble constant), 
not to $0$,  as $y\to \pm y_0$ as shown in Fig.\ref{dSnograviton}.  
The disappearance of the mass gap is peculiar to 
the scalar perturbation for the thick de Sitter brane. 

\subsection{Thick Anti-De Sitter Brane Case}

At last, let us consider the thick AdS brane case. 
In the case of the AdS brane, we decompose 
$F(z, x^{\mu})$ in eq.(\ref{eq:Schrodinger equation}) as 
\begin{equation}
F(z, x^{\mu}) = \int dm f_m(z) g_m(x^{\mu}),
\end{equation}
and we get the following equation for $g_m(x^{\mu})$,
\begin{equation}
\gamma^{\rho\lambda}g_m(x^{\mu})_{|\rho\lambda}
=m^2 g_m(x^{\mu}),
\label{eq:equation_of_g_m}
\end{equation}
where $m$ is the four-dimensional mass.
It is known that eq.(\ref{eq:equation_of_g_m}) can be solved with 
suitable harmonic functions and there is Breitenlohner-Freedman 
bound which allows 
the tachyonic mass to some extent from the condition of the normalization
\cite{AGMOO}-\cite{QFTinAdS}. 
From Breitenlohner-Freedman bound, the mass $m$ is bounded as 
\begin{equation}
m^2 \geq -\frac{9}{4}. 
\end{equation}
It means that even when there are the solution 
with $-\frac{9}{4} \leq m^2 < 0$, 
such solutions are stable in spite of the tachyonic mass. 

\begin{figure}[t]
\begin{center}
\includegraphics[width=8cm]{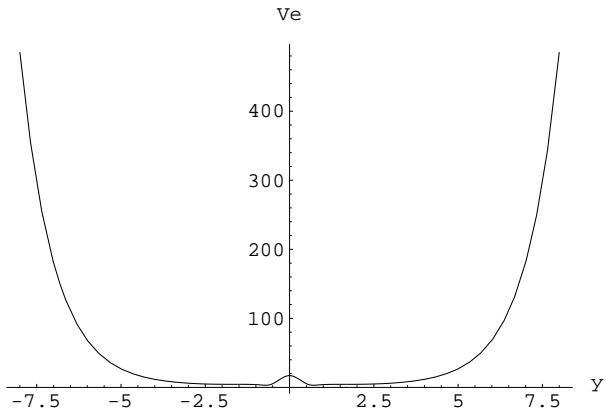}
\caption{The effective potential of the scalar perturbation in the
case of the thick AdS brane. Here we use $n=1$ warp factor of 
(\ref{eq:thick_AdS_warp_factor}) and we set $n=1, y_0=1, \epsilon=0.4$.}
\label{effe_pot_of_AdS_thick-brane}
\end{center}
\end{figure}

Then we examine the effective potential in the case of the thick AdS
brane.
Substituting the warp factor (\ref{eq:thick_AdS_warp_factor}) into 
eq.(\ref{eq:effective potential}) and setting $K=-1$, we can get $V_e$.
We show the numerical result in Fig.\ref{effe_pot_of_AdS_thick-brane}. 

Clearly, the effective potential is positive definite also in this 
case. As the effective potential is positive, 
there is no solution with $m^2 < -\frac{9}{4}$, so we can 
conclude that the thick AdS brane is stable under the scalar
perturbation as well as the thick Poincar\'{e} and the thick de Sitter 
brane. 

Next we pay attention to the shape of the effective potential. 
The upheaval appears around $y=0$ just as in the case of the Poincar\'{e} 
and of the de Sitter brane. 
This upheaval represents that the thick branes behave repulsively.
But the effective potential in the AdS case diverges as $y\to\pm\infty$.
This behavior is unique to the AdS brane system.  
This divergence is caused by the divergence of the warp factor 
at $y\to\pm\infty$. 

Note that we treat the metric which coincides with an exact 
$AdS_5$ only at infinity. 
We may be able to consider the metric which 
coincides an exact $AdS_5$ at finite $y$. But if we use such a metric, 
$\varphi^{\prime}$ becomes $0$ at that point and the effective 
potential appears to diverge there. 
In fact, if we take $\epsilon=0$ in (\ref{eq:thick_AdS_warp_factor}), 
we find that $\varphi^{\prime}=0$ at $y=\pm y_0$ and the effective 
potential diverges there. But note that the warp factor itself 
does not diverge at $y=\pm y_0$. This shows that this divergence has 
a different origin from the divergence at infinity. 
And we can say that the divergence at $y=\pm y_0$ has no physical
meaning. 
In fact, if we take a suitable variable and rewrite the 
effective potential with it, 
we will see no divergence at  $y=\pm y_0$.

\section{Conclusions and Discussions}

In this paper we proposed three types of the thick brane models 
and analyzed the stability of them. 
The three types of the thick branes are the solutions 
of the Einstein equations with non-trivial dilatons and potentials. 
Furthermore, these solutions are non-singular in the whole spacetime 
even at the location of the brane. 

At first, we considered gravitons in these thick brane 
systems. In all cases, the effective potentials of the gravitons 
show that there are bound states which localize around the brane. 
Such gravitons make the four-dimensional gravity on the brane 
if we take the thin brane and low energy limit. 
Here, note that there are ambiguities to estimate 
the corrections to the four-dimensional gravity due to the KK-modes. 
This issue is deserved for a further investigation. 

Next we analyzed the scalar perturbations of the thick brane systems. 
We write the master equations of the scalar perturbations explicitly, 
and we get the effective potentials of the scalar perturbations 
in the three cases, respectively. 
As a result, we see all of the thick branes are stable 
under the scalar perturbations.
This is because the effective potentials of  
the scalar perturbations are positive definite. 
But the shape of the effective potentials of the scalar perturbation 
are very different from that of the gravitons. 
In all cases, there is no bound state because there is 
no hole around the location of the brane in the effective 
potentials of the scalar perturbations. 
Furthermore, there is an upheaval around the 
location of the brane. So we conclude that the three types of 
the thick branes behaves repulsively against the scalar perturbations. 

And the asymptotic behavior of the effective potential of 
the scalar perturbation in the case of the thick de Sitter brane 
is also different from that of the
graviton or the free test scalar field propagating 
in the same background. 
As $y\to\pm y_0$, the effective potential of the graviton approaches 
to $\frac{9}{4}H^2$, not to $0$. 
This mass gap can be seen in various 
analyses on de Sitter brane models\cite{shinpei}. But for the scalar
perturbation in the thick de Sitter brane system, the mass gap
disappeared. That is, the effective potential 
of the scalar perturbation approaches to $0$ 
as $y\to\pm y_0$. This phenomenon is peculiar to the scalar
perturbation including the back reaction. 

And the shape of the effective potential in the case of the thick AdS 
brane is distinctive. 
It diverges as $y\to\pm\infty$, which is caused by 
the divergence of the warp factor at $y=\pm\infty$. In the case of the 
thick Poincar\'{e} brane or of the de Sitter brane, 
the warp factors do not diverge in the whole spacetimes.  

Consequently, we have concluded that all of the three branes 
are stable under the scalar perturbations.  
But we should note that these analysis
have been done classically. 
On the analogy of the two analyses on the 
Schwarzschild black holes (i.e., the analysis by Regge-Wheeler and 
the analysis by Hawking), the thick de Sitter brane 
might be unstable quantum mechanically. 
In fact, as the de Sitter spacetime has a temperature, 
the spacetime may radiate and get to the Poincar\'{e} 
spacetime\cite{Smolin}. 
So we may have to treat this system quantum mechanically. 
We leave this issue for the future work. 

At last we refer the relation between the thickness and the 
non-commutativity. 
As we have mentioned, 
we can make smooth warp factors in various ways. 
For example, to construct the thick Poincar\'{e} brane, 
we can introduce another ``thickness'' parameter $\lambda\theta^2$,  
\begin{equation}
a^2(z) = e^{2\alpha(y(z))} = \frac{1}{e^{2y}+\lambda\theta^2 e^{-2y}}.
\label{eq:warp factor with lambda}
\end{equation}
If we set $\lambda\theta^2$ to $0$, $AdS_5$ spacetime with the 
Poincar\'{e} slicing is recovered.
Here we introduce the new variable $\xi$, which is defined as 
\begin{equation}
e^{-2\xi} \equiv \sqrt{\lambda\theta^2}e^{-2y}.
\end{equation}
Using $\xi$, we can rewrite the warp factor 
(\ref{eq:warp factor with lambda}) into the following form,
\begin{equation}
a^2(\xi) = \frac{1}{\sqrt{\lambda\theta^2}}
\cdot\frac{1}{e^{2\xi}+e^{-2\xi}}.
\label{eq:warp factor with xi}
\end{equation}
So (\ref{eq:warp factor with lambda}) coincides with 
(\ref{eq:warp factor for flat}) with $n=1$ and 
$e^{2y_0} = 1/\sqrt{\lambda\theta^2}$.
From the above equation, 
we can say that (\ref{eq:warp factor with lambda}) is 
one of the warp factors of the thick Poincar\'{e} brane models and that 
$\lambda\theta^2$ is one of the parameters which determine 
the thickness. 

On the other hand, $\lambda\theta^2$ seems to be related to the
non-commutative geometry. From the discussion of the AdS/CFT
correspondence, it is known that there is the classical
solution of the supergravity which corresponds to $\mathcal{N}=4$ 
super Yang-Mills theory in the non-commutative spacetime
\cite{Hashimoto},\cite{MR}. This classical solution is 
given by 
\begin{equation}
ds^2 = dy^2 + \left(\frac{1}{e^{2y}+\lambda\theta^2 e^{-2y}}\right)
\eta_{\mu\nu}dx^{\mu}dx^{\nu} + d\Omega_5^2,
\label{eq:non-com_warp}
\end{equation}
where $d\Omega_5^2$ is the metric of $S^5$ and clearly  
(\ref{eq:warp factor with lambda}) coincides with the warp factor of 
(\ref{eq:non-com_warp}). 

$\mathcal{N}=4$ Super Yang-Mills 
theory in the non-commutative spacetime is realized on the D3-branes 
with the non-zero expectation value of the B-filed. 
In this context, $\lambda$ is 
t'Hooft coupling of $\mathcal{N}=4$ Super Yang-Mills theory in 
the non-commutative spacetime and $\theta$ is the expectation value of 
the B-filed. So we can call $\lambda\theta^2$ the non-commutative 
parameter. 

From above discussion, we can interpret $\lambda\theta^2$ in two ways,  
that is, as the thickness parameter and as the non-commutative parameter. 
So we can expect (\ref{eq:warp factor with lambda}) 
contains some effects due to the non-commutativity. 

There is another thing we should mention here. 
In this paper, we construct the thick de Sitter and the thick 
Anti-de Sitter brane system. 
On the analogy of the case of the thick Poincar\'{e} 
brane, we cannot deny the possibility that the thick de Sitter 
and the thick anti-de Sitter brane system also correspond to
some field theories. 
If so, the corresponding theories may be the quantum 
filed theories in the curved spacetime with the non-commutative 
coordinate. As we have not known any theories in the non-commutative 
curved spacetime, this topic is very interesting. 

Anyway, we want to derive thick brane systems from the ten-dimensional 
supergravity or from the eleven-dimensional M-theory. 
We expect that we might be able to interpret the non-commutativity in 
the context of these theories.   
We leave these themes for future works.


\section*{Acknowledgements}
The work of K.K. was supported by JSPS.



\end{document}